\documentclass[journal=nalefd,manuscript=letter]{achemso}
\setkeys{acs}{articletitle = true}

\usepackage[version=3]{mhchem} 

\usepackage{amssymb}
\usepackage{amsmath}
\usepackage{hyperref}
\usepackage{graphicx}
\usepackage{url} 
\usepackage{enumitem}
\usepackage{mhchem}
\usepackage{gensymb}
\usepackage[export]{adjustbox}
\usepackage{array}
\usepackage{makecell}
\usepackage{listings}
\usepackage{pdfpages}

\newcommand{\onlinecite}[1]{\hspace{-1 ex} \nocite{#1}\citenum{#1}} 

\author{Alison E. Rugar}
\affiliation{E. L. Ginzton Laboratory, Stanford University, Stanford, California 94305, United States}
\email{arugar@stanford.edu}
\author{Haiyu Lu}
\affiliation{Department of Physics, Stanford University, Stanford, California 94305, United States}
\altaffiliation{Geballe Laboratory for Advanced Materials, Stanford University, Stanford, California 94305, United States}
\author{Constantin Dory}
\affiliation{E. L. Ginzton Laboratory, Stanford University, Stanford, California 94305, United States}
\author{Shuo Sun}
\affiliation{E. L. Ginzton Laboratory, Stanford University, Stanford, California 94305, United States}
\author{Patrick J. McQuade}
\affiliation{Department of Materials Science and Engineering, Stanford University, Stanford, California 94305, United States}
\altaffiliation{Stanford Institute for Materials and Energy Sciences, SLAC National Accelerator Laboratory, Menlo Park, California 94025, United States}
\author{Zhi-Xun Shen}
\affiliation{Department of Physics, Stanford University, Stanford, California 94305, United States}
\altaffiliation{Geballe Laboratory for Advanced Materials, Stanford University, Stanford, California 94305, United States}
\altaffiliation{Stanford Institute for Materials and Energy Sciences, SLAC National Accelerator Laboratory, Menlo Park, California 94025, United States}
\author{Nicholas A. Melosh}
\affiliation{Department of Materials Science and Engineering, Stanford University, Stanford, California 94305, United States}
\altaffiliation{Stanford Institute for Materials and Energy Sciences, SLAC National Accelerator Laboratory, Menlo Park, California 94025, United States}
\author{Jelena Vu\v{c}kovi\'c}
\affiliation{E. L. Ginzton Laboratory, Stanford University, Stanford, California 94305, United States}

\title[SnV- delta doping and growth]
  {Generation of Tin-Vacancy Centers in Diamond via Shallow Ion Implantation and Subsequent Diamond Overgrowth}

\abbreviations{IR,NMR,UV}

\begin{document}
\begin{abstract}
Group-IV color centers in diamond have garnered great interest for their potential as optically active solid-state spin qubits. Future utilization of such emitters requires the development of precise site-controlled emitter generation techniques that are compatible with high-quality nanophotonic devices. This task is more challenging for color centers with large group-IV impurity atoms, which are otherwise promising because of their predicted long spin coherence times without a dilution refrigerator. For example, when applied to the negatively charged tin-vacancy (SnV$^-$) center, conventional site-controlled color center generation methods either damage the diamond surface or yield bulk spectra with unexplained features. Here we demonstrate a novel method to generate site-controlled SnV$^-$ centers with clean bulk spectra. We shallowly implant Sn ions through a thin implantation mask and subsequently grow a layer of diamond via chemical vapor deposition. This method can be extended to other color centers and integrated with quantum nanophotonic device fabrication.

\textit{This document is the Accepted Manuscript version of a Published work that appeared in final form in
Nano Letters, copyright © American Chemical Society after peer review and technical editing by the publisher.
To access the final edited and published work see \url{https://pubs.acs.org/articlesonrequest/AOR-HGJZcs32FW9nR3n8mwcT}.}
\end{abstract}


Group-IV color centers in diamond have emerged as promising candidates for optically active, solid-state spin qubits\cite{AharonovichDiamondCCReview2011,Atature2018,Awschalom2018,AharonovichTrusheimReview2019}. These color centers are comprised of a split vacancy in the diamond lattice and an interstitial group-IV atom. The inversion symmetry of this structure provides group-IV color centers beneficial properties such as insensitivity to electric field fluctuations to first order and  high Debye-Waller factors\cite{SipahigilSiVPRL2014}. These color centers also possess long-lived electron spins that can be harnessed as quantum memories\cite{LukinSiV100mK,LoncarSiVStrainCoherence,LukinSiVQuantumRegister2019_shorter}. All of these characteristics make group-IV color centers well suited to interface optical photons in nanophotonic platforms for applications in quantum networks.

An outstanding challenge in implementing these color centers in scalable applications is their generation.
The two most common methods of group-IV color center generation are ion implantation and synthesis. Ion implantation facilitates site-controlled generation of color centers by using either a mask\cite{LukinSiVQuantumRegister2019,EnglundHighAspectRatio_NanoLetters2015} or focused ion beam (FIB)\cite{SchroederFIBSiV2017,ZhouFIBGeV2018}. However, the quality of ion-implanted emitters is often degraded by the large amount of damage introduced during implantation\cite{AharonovichTrusheimReview2019}. Synthesis techniques such as high-pressure high-temperature (HPHT) growth and chemical vapor deposition (CVD) growth often yield higher quality, more stable emitters with lower inhomogeneous broadening than ion implantation\cite{CraneHPHTSiV2019,NeuSiV_Iridium,PalyanovGeV2015,IwasakiGeV2015}. Unfortunately, synthesis techniques do not enable site-controlled generation. 
A better color center generation method is severely lacking.

Here we introduce a novel method of generating color centers that overcomes the trade-off between site control and surface and emitter quality. 
We apply this method, which we call \underline{s}hallow \underline{i}on \underline{i}mplantation and \underline{g}rowth (SIIG), to generate negatively charged tin-vacancy (SnV$^-$) centers. Interest in larger group-IV atoms such as the SnV$^-$ center has grown because the larger spin-orbit interaction enables longer spin coherence times without requiring dilution refrigerators\cite{GaliPRX}. However, the issue of generating color centers without significant lattice damage is all the more pertinent to color centers with larger atoms.
High-quality SnV$^-$ centers have been generated only with HPHT annealing or growth\cite{IwasakiSnV,GoerlitzSnV2019,Ekimov_SnV_HPHT2018}. Of these two methods, only HPHT annealing enables site-controlled color center generation, but it also may damage the diamond surface\cite{HPHTdiamondBook} and thus inhibit subsequent nanophotonic device fabrication.
Ion implantation with vacuum annealing, while compatible with nanophotonic fabrication, produces samples that display bulk photoluminescence (PL) spectra with extraneous peaks\cite{Rugar_SnV_PRB2018,EnglundSnV}. 
With the SIIG method, we demonstrate site-controlled color center generation of SnV$^-$ centers. 
We also demonstrate that this method generates high-quality SnV$^-$ centers consistently---the extraneous emission peaks around 631~nm and 647~nm in the bulk spectra are absent in the SIIG sample. 
This result, to our knowledge, is unprecedented in the absence of HPHT conditions.

\begin{table}[h]
    \centering
    \resizebox{\textwidth}{!}{
    \begin{tabular}{|c|c|c|c|}
    \hline \textbf{Sample} & \textbf{Pre-implantation steps} & \makecell{\textbf{Implantation}\\\textbf{conditions}} & \textbf{Post-implantation steps} \\
    \hline A & tri-acid clean, 500~nm O$_2$ plasma etch & \makecell{370~keV,\\$2\times 10^{13}$ cm$^{-2}$} & \makecell{vacuum anneal\\30~minutes at 800$\degree$C,\\30~minutes at 1100$\degree$C} \\
    \hline B & \makecell{tri-acid clean, 500~nm O$_2$ plasma etch,\\spin on PMMA, e-beam lithography} & \makecell{1~keV,\\$2\times 10^{13}$ cm$^{-2}$} & \makecell{remove PMMA,\\H$_2$ plasma clean,\\90~nm MPCVD diamond growth} \\
    \hline
    \end{tabular}
    \caption{Summary of the fabrication steps performed on each of the samples studied.}
    \label{samplesummary_table}
    }
\end{table}
Two samples, named A and B, summarized in Table \ref{samplesummary_table}, are studied in this paper. 
Both samples started as electronic grade diamond plates from Element Six. The diamond chips were cleaned in a boiling tri-acid solution (1:1:1 sulfuric:nitric:perchloric acids). An oxygen (O$_2$) plasma etch was then used to remove 500~nm of diamond.
Sample A was implanted at CuttingEdge Ions with $^{120}$Sn$^+$ ions at 370~keV and $2\times10^{13}$~cm$^{-2}$. It was then annealed under vacuum ($\sim 10^{-4}$~Torr) at 800$\degree$C for 30~minutes and 1100$\degree$C for 30~minutes. Sample A provides a baseline for color center quality to which we compare the results of the SIIG method. 

\begin{figure}[h]
\includegraphics[width=0.5\textwidth,]{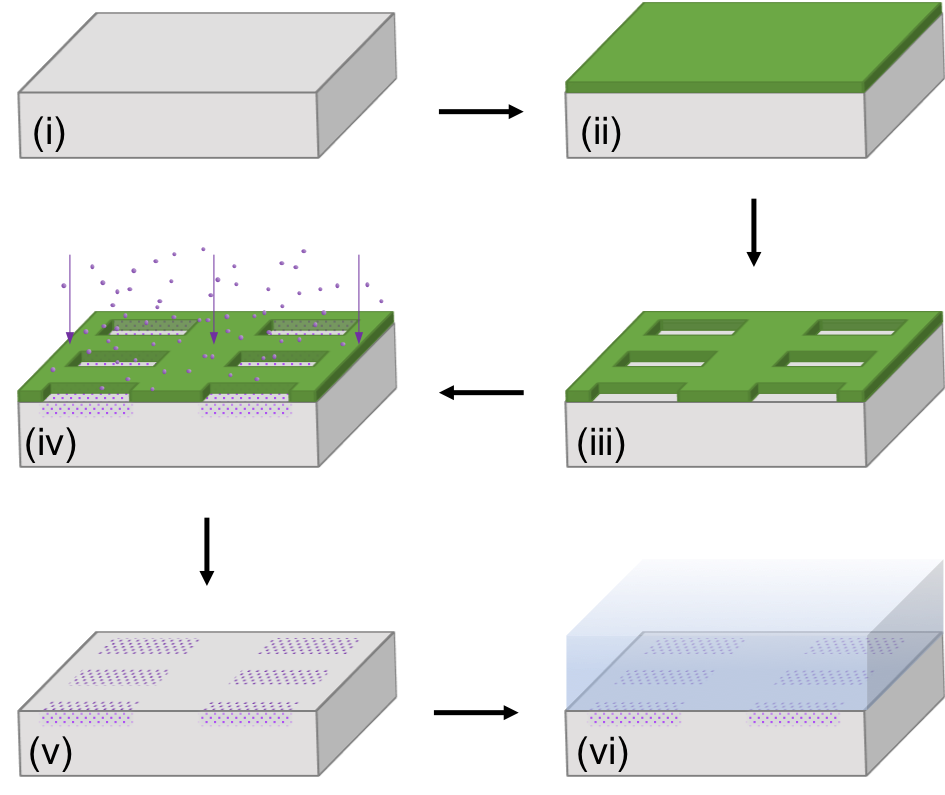}
\caption{Schematic of the SIIG method of SnV$^-$ center generation used in this paper. (i) Starting with electronic grade diamond (gray), perform tri-acid clean and etch $\sim500$~nm with O$_2$ plasma; (ii) spin-coat $\sim50$~nm of PMMA (green); (iii) pattern the PMMA via e-beam lithography; (iv) perform 1~keV implantation of $^{120}$Sn$^+$ ions (purple); (v) remove PMMA with Remover PG; (vi) clean surface with H$_2$ plasma and immediately grow 90~nm of diamond via MPCVD.}
\label{SIIGmethod_fig}
\end{figure}

Sample B was prepared via the SIIG method with a patterned implantation mask, as shown schematically in Figure \ref{SIIGmethod_fig}.  Sample B was spin-coated with $\sim50$~nm of poly(methyl methacrylate) (PMMA, specifically 950 PMMA A2). The PMMA was then patterned via electron-beam (e-beam) lithography in arrays of square holes ranging in side length from 20~nm to 150~nm. Several 10~$\mu$m$\times$10~$\mu$m holes were also patterned for bulk measurements. After the PMMA was developed in a $\sim5\degree$C 3:1 solution of isopropanol:water, sample B was sent to CuttingEdge Ions for shallow ion implantation $^{120}$Sn$^+$ ions at 1~keV with a dose of $2\times10^{13}$~cm$^{-2}$. The PMMA mask was then removed with Remover PG, and we began the diamond growth procedure. The surface was cleaned with hydrogen (H$_2$) plasma. Immediately thereafter we grew a nominally 90-nm thick layer of diamond via microwave-plasma CVD (MPCVD, Seki Diamond Systems SDS 5010) with 300~sccm H$_2$; 0.5~sccm CH$_4$; stage temperature of 650$\degree$C; microwave power of 1100~W; pressure of 23~Torr. 

By first implanting with low energy, we localize ions and related lattice damage to within $\sim2$~nm of the surface, with lateral and longitudinal straggles of 3~\AA, as calculated in Stopping and Range of Ions in Matter (SRIM) simulations. The H$_2$ plasma then removes unwanted sp$^2$-bonded carbon that may have resulted from implantation damage. Similarly, during the MPCVD growth, sp$^2$ bonds are removed and new sp$^3$ bonds are formed, healing and reconstructing the damaged lattice near the surface while incorporating the Sn atoms into the growing diamond lattice to form SnV$^-$ centers. As an added benefit, MPCVD growth enables us to control the depth of the $\delta$-doped layer of Sn ions. We are thus freed of limitations imposed by the maximum implantation energy available and the increased ion straggle expected from higher-energy implantation.

\begin{figure}[h]
\includegraphics[width=0.5\textwidth,]{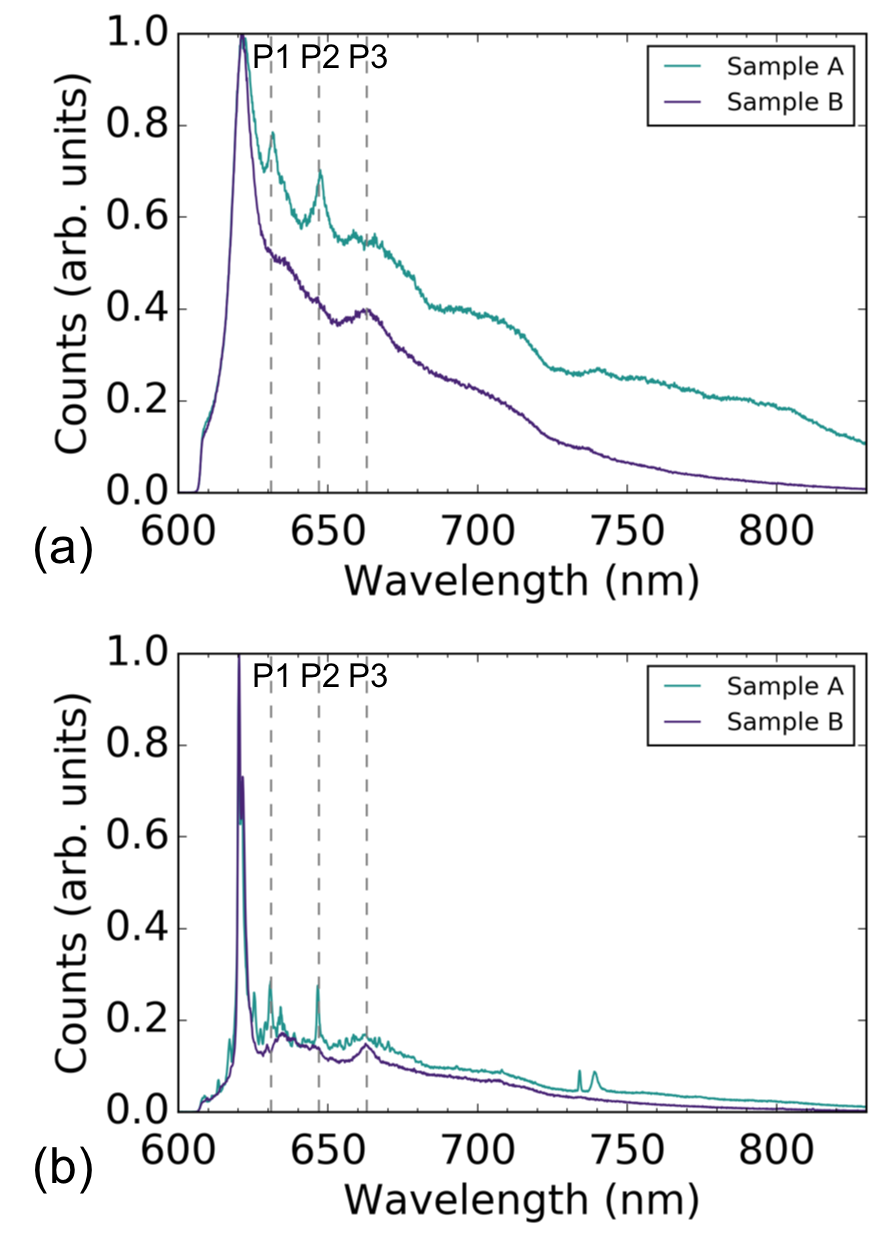} 
\caption{PL spectra from two methods. (a) PL spectra at room temperature for samples A (teal) and B (purple). Gray dashed lines, P1, P2, and P3, respectively mark 631~nm, 647~nm, and 663~nm. (b) Same as (a) for samples held at 5~K. Features around 631~nm and 647~nm are present only in the spectra for sample A.}
\label{compare_PL_spectra_fig}
\end{figure}

We first compare the average bulk PL spectra of samples A and B acquired at room temperature and 5~K, shown in Figure \ref{compare_PL_spectra_fig}.
Although the spectra for both samples display a prominent SnV$^-$ zero-phonon line (ZPL) around 620~nm, they differ significantly in the rest of the measured wavelength range. We note three wavelengths around which the two spectra differ: 631~nm, 647~nm, and 663~nm, marked in Figure \ref{compare_PL_spectra_fig} as P1, P2, and P3 respectively. 

Around P1 and P2, peaks are present in the spectra of sample A but not sample B. These peaks have been observed in previous studies of Sn-implanted diamond\cite{IwasakiSnV,TchernijSnV}. The suppression of these two peaks as observed in sample B has been previously achieved only through HPHT processing\cite{IwasakiSnV,Ekimov_SnV_HPHT2018}. Because sample B never underwent any HPHT process, the SIIG method to form SnV$^-$ centers is unique from simply annealing an implanted sample.

The origins of the peaks at P1 and P2 are, as yet, unclear.
Some argue that because these peaks have been found to disappear under certain conditions, they are not related to the SnV$^-$ center\cite{Ekimov_SnV_HPHT2018}. Under this interpretation, the absence of these peaks in the PL spectrum of sample B indicates that the SIIG method enables the formation of SnV$^-$ centers and mitigates the creation of the defects responsible for the 631- and 647-nm emission. Another interpretation of the 631-nm emission is that it is a quasi-local mode in the phonon sideband of the SnV$^-$ center which is allowed only when the symmetry of the defect center is broken, \textit{e.g.}, via strain\cite{TchernijSnV,GoerlitzSnV2019}. Following the argument in Ref. \onlinecite{GoerlitzSnV2019}, the disappearance of the 631-nm line in the bulk spectrum for the SIIG method would imply that the SIIG method produces a low-strain, high-quality crystal around the SnV$^-$ centers.


While we currently cannot definitively state the source of the peak at P1, we have investigated in greater detail the origin of the peak at P2.
The 647-nm peak, at P2, was previously observed by Iwasaki \textit{et al.}\cite{IwasakiSnV} and Tchernij \textit{et al.}\cite{TchernijSnV}. In the previous studies the 647-nm emission was observed only in Sn-implanted regions\cite{TchernijSnV} and was found to vary in intensity depending on the annealing temperature, ultimately disappearing after a HPHT anneal at 2100$\degree$C and 7.7~GPa\cite{IwasakiSnV}. These observations indicate that the 647-nm peak may be from a Sn-related color center that is not the SnV$^-$ center. In the Supporting Information\cite{supportinginfo}, we present room-temperature and cryogenic spectra exhibiting the 647-nm line when sample A was excited with 632.8-nm laser light, further evidence that the 647-nm peak is not from the SnV$^-$ center.

The third feature around 663~nm presents in sample A as a broad bump, while it appears in sample B as a narrower peak. The presence of a peak around 663~nm is consistent with not only the experimental results of Iwasaki \textit{et al.} after a HPHT anneal\cite{IwasakiSnV} and SnV$^-$ centers synthesized via HPHT\cite{Ekimov_SnV_HPHT2018} but also density functional theory (DFT) calculations of the SnV$^-$ spectrum\cite{GaliPRX}. Upon closer inspection of the cryogenic spectrum shown in Figure \ref{compare_PL_spectra_fig}(b), we find that the phonon sideband of sample B holds quite a good qualitative resemblance to that calculated by DFT in Ref. \onlinecite{GaliPRX}. 
For example, in addition to the peak around 663~nm, the spectrum of sample B shares with the DFT-predicted spectrum peaks around 635~nm and 645~nm.
These similarities to DFT calculations indicate that the emission spectrum of sample B is dominated by SnV$^-$ emission. By contrast, the SnV$^-$ PL spectrum of sample A seems to be obscured by additional emission.

\begin{figure*}[h]
\includegraphics[width=1\textwidth,]{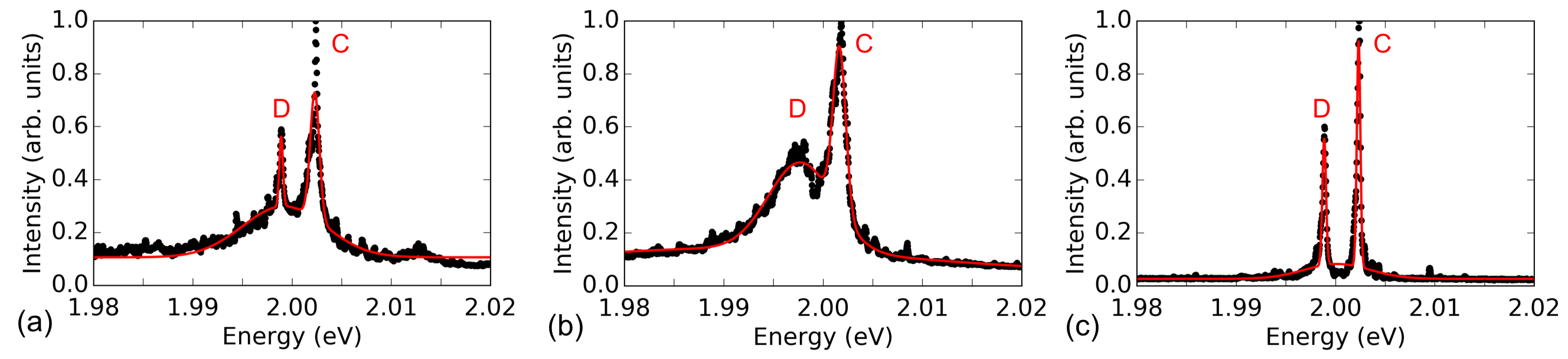} 
\caption{Inhomogeneous broadening. (a) Spectrum of the C and D transitions at 5~K for sample A. Spectrum is an average of four spectra collected via multimode fiber at different spots on the sample. Fit to data (black dots) is shown in red. (b) Same as (a) for sample B. (c) Same as (a) for an average of spectra collected via multimode fiber at three different spots on a third sample that was made with SIIG with a different initial etch depth and a lower implantation dose.}
\label{compare_inhomogeneousBroadening_fig}
\end{figure*}

Another figure of merit that can be used to characterize emitter quality is the inhomogeneous broadening. For group-IV color centers, inhomogeneous broadening typically comes from variations in strain throughout the sample\cite{EvansSiVPRApplied2016,LukinSiVsCoupled} and can therefore serve as an indicator of sample quality after emitter generation. We quantify the inhomogeneous broadening for both samples A and B by averaging PL spectra acquired at four different spots on the sample and fitting Gaussians to the C and D ZPL transitions in the resulting spectra, as shown in Figure \ref{compare_inhomogeneousBroadening_fig}. To demonstrate how sample preparation can improve inhomogeneous broadening in SIIG samples, a third sample, sample C, is also studied. Sample C was prepared via SIIG, but had a deeper initial etch and lower implantation dose than sample B. More details on the fits and the preparation of sample C can be found in the Supporting Information\cite{supportinginfo}.

For sample A, the fit to the average ensemble PL reveals full widths at half-maxima (FWHMs) of $263\pm5$~GHz and $105\pm5$~GHz for the C and D transitions respectively at 5~K. In sample B, the FWHMs for the average ensemble C and D transitions are broader at $363\pm4$~GHz and $1858\pm20$~GHz respectively. By contrast, for sample C, the FWHMs in the average ensemble PL are much narrower than those of sample B and comparable to or narrower than those of sample A: $101\pm1$~GHz and $105\pm2$~GHz for the C and D transitions respectively. The stark contrast between the broadening observed in samples B and C indicates the importance of sample preparation with the SIIG method and the potential to further reduce the inhomogeneous broadening by optimizing the sample preparation. The narrow distribution of emitters in sample C is promising because it is narrower than that in the implanted and vacuum annealed sample A. Furthermore, the $\sim100$~GHz inhomogeneous broadening displayed by sample C is narrow enough to be overcome by techniques such as Raman tuning, which has a demonstrated tuning range of 100~GHz for a silicon-vacancy (SiV$^-$) center\cite{SipahigilScience2016,SunRamanPRL}. 

\begin{figure*}[]
\includegraphics[width=1\textwidth,]{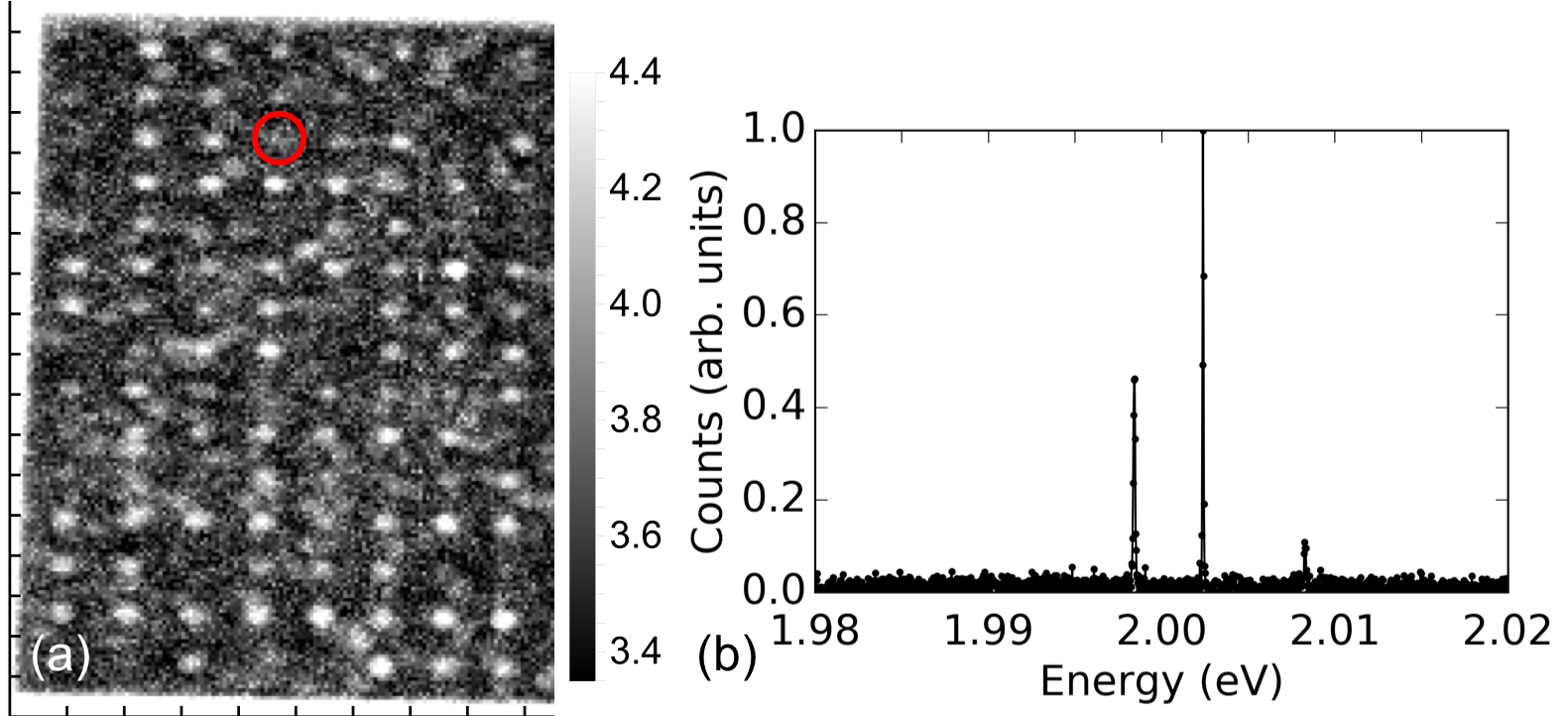}
\caption{Site-controlled generation of SnV$^-$ centers. (a) PL map resulting from color center generation with a 15-row$\times$8-column array of 30-nm$\times$30-nm holes. The color scale bar is $\log_{10}$(counts/s). Ticks on axes are in 1-$\mu$m increments. (b) PL spectrum of SnV$^-$ center located at site circled in red in (a).}
\label{patterned_implantation_fig}
\end{figure*}

A potential application of the SIIG method of generating SnV$^-$ centers is patterned implantation. 
By requiring very low ion implantation energy, this method obviates the need for extremely thick implantation masks which limit the minimum achievable hole size\cite{LukinSiVQuantumRegister2019} or high aspect ratio masks\cite{EnglundHighAspectRatio_NanoLetters2015} which are quite complicated to fabricate and the efficacy of which is ultimately limited by lateral ion straggle. Instead, we are able to use a $\sim50$-nm thick layer of PMMA to stop the implanted ions. We patterned arrays of square holes in the PMMA via e-beam lithography, as described above. 
In Figure \ref{patterned_implantation_fig} we focus on a region that was masked with an array of 30-nm holes. 
$78\%$ of the holes in the region yielded at least one---often more---SnV$^-$ center, resulting in an array of SnV$^-$ centers apparent in the PL map of Figure \ref{patterned_implantation_fig}(a). From this array, we selected one of the sites, circled in red, that has a low number of SnV$^-$ centers. The PL spectrum measured at 5~K for this site is presented in Figure
\ref{patterned_implantation_fig}(b). With this array, we estimate the conversion efficiency to be at least 1$\%$ (see Supporting Information for details\cite{supportinginfo}).

We have introduced a new method for generating SnV$^-$ centers that yields clean, consistent bulk PL spectra without the need for HPHT processing and that enables high-precision placement of color centers with a simple mask of PMMA. The use of low-energy ion implantation in the SIIG method significantly reduces the extent of lattice damage, length of ion straggle, and thickness of PMMA required to stop ions. These improvements are key to precisely placing high-quality color centers in arrays. The subsequent MPCVD growth of diamond facilitates SnV$^-$ formation and can be continued arbitrarily to position the $\delta$-doped layer of color centers at the desired depth. 
The SIIG method may thus be incorporated with existing diamond fabrication techniques\cite{BurekHighQSiVCavity2014,Barclay_QuasiIsotropicEtch2015,Mouradian1DPhC2017,Wan2DPhC2018,MitchellMicrodisksAPLPhotonics} to precisely place color centers within diamond nanostructures\cite{LukinSiVQuantumRegister2019,DoryOptimizedDiamondPhotonics} for applications in nanophotonics and phononics.

We find that extraneous spectral features typically observed in Sn-implanted samples in the absence of HPHT processing disappear with the SIIG method. This observation indicates that the SIIG method is unique from standard implantation and annealing techniques. The SIIG method is also distinct from conventional CVD-based color center generation\cite{NeuSiV_Iridium,IwasakiGeV2015} because the partial pressure of Sn in the gas phase is minimal. We are thus able to generate SnV$^-$ centers with well-known plasma parameters typically used to grow pure diamond. The compatibility of SIIG with the standard growth conditions of the host material speaks to the versatility of the method. We believe SIIG can be extended to other color centers in not only diamond but a multitude of other host materials\cite{Awschalom2018} and may be of great use in new color center\cite{EnglundPbV,TchernijPbV,Harris_GroupIIIcolorcenters_2019} discovery.

\begin{acknowledgement}

This work is financially supported by Army Research Office (ARO) (award no. W911NF-13-1-0309); National Science Foundation (NSF) RAISE TAQS (award no. 1838976); Air Force Office of Scientific Research (AFOSR) DURIP (award no. FA9550-16-1-0223); Department of Energy, Basic Energy Sciences (BES) - Materials Science and Engineering; and SLAC LDRD. A.E.R. acknowledges support from the National Defense Science and Engineering Graduate (NDSEG) Fellowship Program, sponsored by the Air Force Research Laboratory (AFRL), the Office of Naval Research (ONR) and the Army Research Office (ARO). C.D. acknowledges support from the Andreas Bechtolsheim Stanford Graduate Fellowship and the Microsoft Research PhD Fellowship. Part of this work was performed at the Stanford Nanofabrication Facility (SNF) and the Stanford Nano Shared Facilities (SNSF), supported by the National Science Foundation under award ECCS-1542152.

\end{acknowledgement}

\begin{suppinfo}

Experimental details and further discussions of inhomogeneous broadening, sample preparation, and conversion efficiency.

\end{suppinfo}



\providecommand{\latin}[1]{#1}
\makeatletter
\providecommand{\doi}
  {\begingroup\let\do\@makeother\dospecials
  \catcode`\{=1 \catcode`\}=2 \doi@aux}
\providecommand{\doi@aux}[1]{\endgroup\texttt{#1}}
\makeatother
\providecommand*\mcitethebibliography{\thebibliography}
\csname @ifundefined\endcsname{endmcitethebibliography}
  {\let\endmcitethebibliography\endthebibliography}{}


\begin{tocentry}
\begin{center}
\includegraphics[width=1\textwidth]{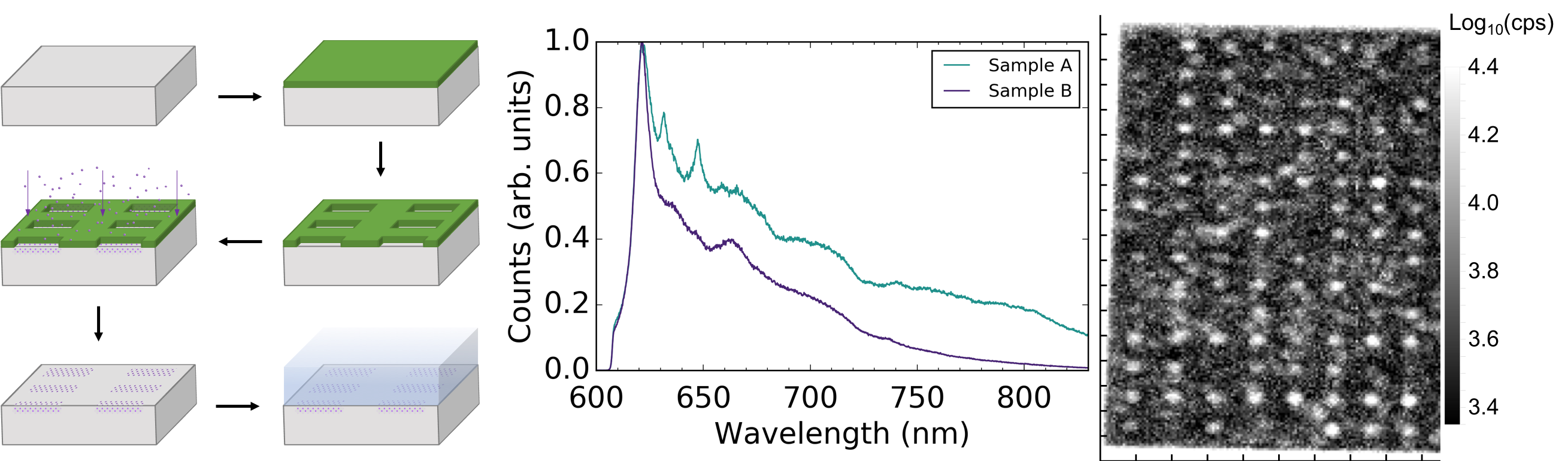}
\end{center}





\end{tocentry}

\end{document}